\documentclass[preprint,aps,floats,tightenlines,eqsecnum, nofootinbib]{revtex4}

\newcommand{\cof}[3]{#1_{#2}^{(#3)}}
\newcommand{\cofs}[3]{#1_{#2}^{*(#3)}}
\newcommand{\var}[2]{\frac{ \delta #1}{ \delta #2}}
\newcommand{\at}[1]{\Big |_{#1}}

\def\ginv{\frac{1}{\sqrt{g}} }
\def\be{\begin{equation}}
\def\ee{\end{equation}}
\def\bea{\begin{eqnarray}}
\def\eea{\end{eqnarray}}

\def\iiv{\int d^4x \;}

\def\dota{a \frac{d}{da}}

\def\lmb{\lambda}

\usepackage[dvips]{graphics}
\usepackage{epsf}
\usepackage{amsfonts}

\begin{document}

\thispagestyle{empty}

\preprint{
\vbox{\hbox{\today} }}

\vspace*{1cm}
\title{Running with the Radius in RS1}
\author{Adam Lewandowski, Michael J. May, Raman Sundrum}
\affiliation{Department of Physics and Astronomy, The Johns Hopkins 
University \\
  3400 North Charles Street,  Baltimore, Maryland 21218}
\setcounter{page}{0}
\begin{abstract}
We derive a 
renormalization group formalism  for the Randall-Sundrum scenario,
where the renormalization scale is set by a floating 
 compactification radius. While inspired by the 
AdS/CFT  conjecture, our results are  derived concretely within 
higher-dimensional effective field theory. 
Matching theories with different radii leads to running hidden brane 
couplings. The hidden brane Lagrangian consists of four-dimensional 
local operators  constructed from the induced value
of the bulk fields on the brane. We find hidden Lagrangians which are 
non-trivial fixed points of the RG flow. 
Calculations in RS1 can be greatly simplified by ``running down''
the effective theory to a small radius.  We demonstrate these simplifications
by studying the Goldberger-Wise stabilization mechanism.  
In this paper, we focus on the classical and tree-level quantum field theory 
of bulk scalar fields, which demonstrates the essential features of the RG 
in the simplest context.
\end{abstract}

\maketitle

\section{Introduction}

Physics in warped higher dimensional spacetimes may play an 
important role in nature, as illustrated by the 
RS1 model~\cite{Randall:1999ee,Randall:1999vf}. However,
effective 
field theory calculations are particularly complicated in such contexts. 
When warp factors vary greatly in the extra dimension(s) it becomes
difficult 
to power-count the size of Feynman diagrams.
 For examples of such calculations see Refs.~\cite{adshard}.  
In 
this paper, we begin a program of honing ``holographic 
renormalization group'' ideas~\cite{holography}, applied to the RS scenario 
in~\cite{rsholography} ,  into a tool for simplifying and gaining
insight 
into warped effective field theory. In particular we will apply this
tool 
to understand the robustness of the Goldberger-Wise stabilization
mechanism~\cite{Goldberger:1999wh}
within effective field theory, although the methods are clearly very
general. We use the AdS/CFT conjecture to inspire the form of a 
concrete and explicit renormalization group procedure, but do not rely 
on any unproven aspects of the conjecture.

The form our RG will take is as a flow of effective field theories in
RS1-type
geometries of varying radius, $r_c$, theories with smaller $r_c$
having 
lower warped down effective energy cutoffs. Two effective theories
lying on 
the same flow are matched to describe the same physics at energies
below 
the lower of the two associated cutoffs. In particular this means that 
matched theories with different radii have the same ``IR'' or ``visible'' 
brane positions, but different ``UV'' or ``hidden'' brane positions.
The matching is accomplished by 
tuning the hidden brane localized
Lagrangian 
as a function of $r_c$. This 4D Lagrangian depends on the values of bulk
fields 
induced on the hidden brane.  We will show that formally the flow of 
hidden brane Lagrangians looks like a Wilsonian RG flow of purely 4D
effective field theories. A crucial feature of the associated 4D effective
Lagrangians is that they are local which greatly simplifies power-counting.  
We identify a conformal fixed point 
in the RG as the 
concrete realization of the AdS/CFT connection within our formalism.
Refs. \cite{skenderis} have developed a related RG formalism focusing on
renormalization of boundary divergences and properties of the holographic
stress tensor.
RG flows associated to branes in codimension two were 
studied in Refs. \cite{wiseflow,Milton}.
While this paper was being
completed, Refs. \cite{deconstruction,falkowski} appeared, which also 
discusses 
similar RG ideas.  

The utility of our RG formalism is this; 
 one can integrate the flow to
take
low-energy calculations in  
a highly warped compactification where they are difficult, to simpler 
calculations 
in a much smaller compactification (smaller $r_c$) where the effect of 
warping is much less significant. Of course this requires computing the 
effective couplings on the hidden brane in the new compactification, but 
this is facilitated by perturbing about the conformal fixed point and by 
working systematically in the small couplings of the theory.

In this paper, we deal only with classical and tree-level quantum 
effective 
field theory involving a bulk scalar field, deferring a treatment of
full effective 
quantum field theory with varying spins. Already at this level, many of the 
non-trivial features of the RG procedure come into play. The paper is 
organized as follows. In Section II, we discuss free field theory 
and point out the basic RG structure and fixed
point.
In Section III, we apply our results to rederive in a simple way the
Goldberger-Wise 
stabilization mechanism when the bulk scalar is identified with 
the one introduced by Goldberger and Wise. (Since we do not make gravity 
dynamical, our analysis is necessarily restricted to neglect the
backreaction 
of the scalar on the geometry, the same approximation made in the
original 
analysis.) In Section IV, we generalize our RG procedure to include a
large 
class of interactions, in the bulk and on branes.  We then use this in 
Section V to 
demonstrate the robustness of the Goldberger-Wise mechanism within
classical 
effective field theory in a manner that lends itself to generalization 
to the quantum case. In Section VI, we switch from the language of
classical 
field theory to consider tree-level Feynman diagrams, setting the stage
for
the case of quantum loops, and showing how one deals with 
general effective interactions. Section VII contains some 
concluding remarks.

\section{Free Theory}
\label{free}

We consider a bulk 5D scalar field, $\chi$, propagating on a 
fixed RS1 background with metric,
\begin{equation}~\label{metric1}
ds^2 = \frac{1}{(kz)^2}(\eta_{\mu \nu} dx^{\mu} dx^{\nu}-dz^2).
\end{equation}
The $z$ coordinate parametrizes the direction in a compact orbifolded
space.  
This is a slice of AdS space bounded by the 
hidden brane at $z=z_h$ and the visible brane at $z=z_v$.  We take 
$1/z_h = k \sim {\cal O}(M_{Plank})$.

In this section we  begin by studying a simple action which is exactly 
quadratic in $\chi$,
\begin{equation}
S = S_{bulk} + S_{vis} + S_{hid},
\end{equation}
where
\begin{eqnarray}
S_{bulk} & = &  \int d^4 x dz \sqrt{G}
\left(\frac{1}{2}(G^{MN} \partial_M
\chi \partial_N \chi) - \frac{1}{2} m^2 \chi^2 \right) \nonumber \\
S_{vis} & = & - \int d^4 x \sqrt{g_{v}} \left( 
\frac{1}{2} \chi \, k  \lambda_{v2} \,
\chi \right) \nonumber \\
S_{hid} & = & - \int d^4 x \sqrt{g_h} \left(\frac{1}{2}\chi \, k
\lambda_{h2} 
 \, \chi \right).
\end{eqnarray}
The brane actions are written in terms of the induced brane metrics,
\begin{eqnarray}~\label{gident}
g^{v}_{\mu \nu}(x) &\equiv& G_{\mu \nu}(x, z=z_v), \nonumber \\
g^{h}_{\mu \nu}(x) &\equiv& G_{\mu \nu}(x, z=z_h),
\end{eqnarray}
and the dimensionless $x$-derivative expansions (the 4D d'Alembertian is $g^{\mu \nu}_h \partial_{\mu} \partial_{\nu}$ for the metric defined in (\ref{metric1}) and (\ref{gident})).
\begin{equation} \label{lambdadefn}
\lambda_{h2} = \lambda_{h2}^{(0)} + k^{-2}\lambda_{h2}^{(2)} g_h^{\mu
\nu} 
\partial_{\mu} \partial_{\nu} + k^{-4}\lambda_{h2}^{(4)} g_h^{\mu \nu} 
g_h^{\rho \sigma} \partial_\mu \partial_\nu \partial_\rho
\partial_\sigma + 
\ldots
\end{equation}
and similarly for $\lambda_{v2}$. The subscript ``$2$'' denotes that the brane
terms are
quadratic in fields, which we generalize later.
We denote the dimensions explicitly in units of the warping constant,
$k$.
Note that 4D general covariance ensures that in the space of
4-momenta, $q$,  the operator $\lmb_{h2} \rightarrow
\lambda_{h2}(q^2z^2_h)$.

The equation of motion is
\begin{equation} \label{bulkeom}
\left( \partial_M \sqrt{G} G^{MN} \partial_N + \sqrt{G} m^2 + 
\sqrt{g_v} k \lambda_{v2} \delta(z-z_v) + 
\sqrt{g_h} k \lambda_{h2} \delta(z-z_h) \right) \chi(x,z) = 0.
\end{equation}
In 4-momentum space this is 
\begin{equation}
\frac{1}{(kz)^3} \left( -q^2 - \partial_z^2 + \frac{3}{z}\partial_z + 
\frac{m^2}{(kz)^2} + \frac{\lambda_{v2}}{z_v} \delta(z-z_v) + 
\frac{\lambda_{h2}}{z_h} \delta(z-z_h) \right) \chi_q(z) = 0
\end{equation}
Strictly in the bulk, $\chi_q(z)$  satisfies the $z_h$ independent 
differential equation
\begin{equation} \label{zhindpeom}
\frac{1}{(kz)^3} \left( -q^2 - \partial_z^2 + \frac{3}{z}\partial_z + 
\frac{m^2}{(kz)^2} \right) \chi_q(z) = 0, 
\end{equation}
with boundary conditions
\begin{eqnarray} \label{bcz}
z \partial_z \chi_q(z) \bigg|_{z=z_h} & = & \frac{\lambda_{h2}}{2}
\chi_q(z_h) 
\\ 
z \partial_z \chi_q(z) \bigg|_{z=z_v} & = & -\frac{\lambda_{v2}}{2} 
\chi_q(z_v).\end{eqnarray}
We now define an ``effective'' hidden brane whose location is a
variable,
$a \in [z_h,z_v)$. 
 We seek a new set of ``effective couplings'' on this brane replacing 
$\lmb_{h2}$ on the original hidden brane,
\begin{equation} 
\lambda_{2}(a) = \lambda_{2}^{(0)}(a) + k^{-2}\lambda_{2}^{(2)}(a) 
g_a^{\mu \nu} 
\partial_{\mu} \partial_{\nu} + k^{-4}\lambda_{2}^{(4)}(a) g_a^{\mu \nu} 
g_a^{\rho \sigma} \partial_\mu \partial_\nu \partial_\rho
\partial_\sigma + 
\ldots ,
\end{equation}
where $g_a^{\mu \nu}$ is the induced metric on the brane,
which imply a new set of boundary conditions at $a$,
\begin{equation} \label{bca}
z \partial_z \chi_q(z) \bigg|_{z=a} = 
\frac{\lambda_2 \big(q^2a^2, a \big) }{2} 
\chi_q(a). 
\end{equation}
We require:
\begin{enumerate}
\item The matching condition that $\chi(x,z)$ which are solutions to the 
original equations of motion (inluding boundary conditions), 
(\ref{bulkeom}), are also solutions to the effective equations of motion
when 
confined to the effective domain $z \in [a,z_v)$.  
\item The boundary condition that $\lambda_{2h}(q^2z_h^2) = 
\lambda_2(q^2a^2,a)$ 
for $a=z_h$. This is obviously consistent with the previous requirement.
\\
\end{enumerate}

We interpret $\lambda_2$ as a set of running couplings, 
$\cof{\lambda}{2}{m}(a)$,  and associated 4D generally-covariant derivative 
operators, $(qa)^m$, just as in (\ref{lambdadefn}).  We would like to
find the 
``renormalization group'' 
flow, $\dota \lambda_2(a)$, of this set.  We proceed by taking the
logarithmic 
derivative of the boundary condition (\ref{bca}), use the bulk equation
of 
motion (\ref{zhindpeom}) as $ z \rightarrow a_+ $ to eliminate 
second $z$-derivatives of $\chi$ at $a$,   
and (\ref{bca}) again to eliminate any first $z$-derivatives of $\chi$
at $a$,
thereby obtaining
\nonumber \\ 
\be \label{lambda2running}
\dota \lambda_2 \big(q^2a^2, a \big) = 4 \lambda_2 -
\frac{\lambda_2^2}{2}+ 
2\frac{m^2}{k^2}-2(qa)^2.
\ee
Clearly, the $a$ dependence in the first variable of 
$\lambda_2(qa, a)$ is entirely fixed 
by 4D general covariance. We will focus on the non-trivial dependence in
the 
second variable, getting a set of $\beta$ functions,
\begin{equation}
\sum_{j} (qa)^j \frac{\partial}{\partial a}  \cof{\lambda}{2}{j} = 
-qa \frac{\partial 
\lambda_2}{\partial (qa)} +  4 \lambda_2 - \frac{\lambda_2^2}{2}+ 2 
\frac{m^2}{k^2}-2(qa)^2.
\end{equation}
Since $q$ varies freely, this represents a set of coupled differential 
equations for the $\lambda_2^{(j)}(a)$.

This flow has a set of fixed points.  
In the vicinity of such fixed points, the 
RG flow simplifies. When $\dota \cof{\lambda}{2}{j}=0$ 
for all $j$ we solve the 
differential equation
\begin{equation}
 -qa \frac{\partial \lmb_2}{\partial (qa)} +  4 \lambda_2 - 
\frac{\lambda_2^2}{2}+ 2 \frac{m^2}{k^2}-2(qa)^2 = 0
\end{equation}
to find the fixed point 
\begin{equation} \label{closedfixpt}
\lambda^*_2 =  (4-2 \nu) + 2 q a  \frac{J_{\nu-1}(qa)}{J_{\nu}(qa)},
\qquad \nu=\sqrt{4 + \frac{m^2}{k^2}}.
\end{equation}
This can be expanded in powers of $q^2$ and therefore corresponds to a 
local fixed-point effective brane Lagrangian when written in position
space. 
There is another solution, involving $N_\nu$'s, 
which is non-analytic in $q$ and therefore cannot be interpreted as a  
 local Lagrangian on the effective brane.  It is therefore rejected. 

 For small momentum ($qa \ll 1$) $\lambda^*_2 \simeq 
\lambda_2^{*(0)}$.  As $q \rightarrow 0$ (\ref{closedfixpt}) tells us 
$\cofs{\lmb}{2}{0}=2 (2+\nu)$. The same result is also obvious from 
 the $\beta$ function equation (\ref{lambda2running}) by first setting 
$q \rightarrow 0$. In fact one can solve (\ref{lambda2running}) for the 
fixed point by 
working order by order in powers of $qa$.  This will generate 
(\ref{closedfixpt}) as a series in $qa$. At leading order
(\ref{lambda2running}) 
implies
\be
 \frac{1}{2} \Big(\lambda_2^{*(0)} \Big)^2 - 4 \lambda_2^{*(0)} - 
2\frac{m^2}{k^2}=0.
\ee
Taking the positive root to match (\ref{closedfixpt}) as $q \rightarrow
0$, we 
again find
\begin{equation} \label{lambda20fixpt}
\lambda_2^{*(0)}=2(2+\nu).
\end{equation}
At higher order in $qa$ we find $\lambda_2^{*(j)} \sim {\cal O}(1)$.  This
method of 
working out the  $\beta$ functions in powers of $qa$ will be 
more useful when we 
add interactions and the coupled RG equations are not soluable in closed
form.

Notice that nowhere did the matching procedure specifically require 
that the bulk metric be AdS space.  For example, 
expressions equivalent to (\ref{lambda2running}) and
(\ref{lambda20fixpt}) 
 can 
be obtained for a 5D flat spacetime with the extra dimension taken as an 
orbifolded circle. Since there is no curvature $k$, we 
measure the coupling $\lambda_2$ with an arbitrary mass scale $\mu$. We
use 
the 
metric 
\begin{equation}
ds^2= \eta_{\alpha \beta} dx^\alpha dx^\beta -\frac{da^2}{(\mu a)^2}
\end{equation}
so that the $\beta$ function has no explicit $a$ dependence.  We find
\begin{equation} \label{minkrunning}
a \frac{d}{da} \lambda_{2 Mink} =  - \frac{1}{2}\lambda^2_{2 Mink} + 
2\frac{m^2}{\mu^2}-2\frac{q^2}{\mu^2}.
\end{equation}
There is no fixed point for $q^2>m^2$.  For $q^2 < m^2$ the fixed point
is
\begin{equation}
\lambda_{2 Mink}^* = \frac{2 m}{\mu} \sqrt{1-\frac{q^2}{m^2}}.
\end{equation}
Thus for massless or sufficiently light bulk fields there is no fixed
point 
about which the RG flow simplifies. In particular there is no guarantee
that 
one can match the original theory to an effective 
one in which the extra dimension is much smaller, with only small
perturbations
to the effective hidden brane action. That is a special property of 
having (nearly) AdS bulk geometry.
 Carefully taking the flat space limit of RS1, one can check that 
(\ref{lambda2running}) makes a smooth transition to (\ref{minkrunning}). 

\section{Goldberger-Wise Radius Stabilization}
\label{stab}

Adding a tadpole term to both branes will allow us to reproduce the 
Goldberger-Wise stabilization mechanism \cite{Goldberger:1999wh} using 
the RG analysis.  Since in
this 
paper we are keeping a fixed background geometry for simplicity, we will 
necessarily be making the same approximation as Goldberger and Wise, of 
neglecting the back-reaction of the scalar profile on the geometry.
However, we expect that the 
generalization to dynamical gravity is straightforward. 
Exact solutions that exhibit stability have been examined 
in~\cite{DeWolfe:1999cp}.  A perturbative approach to stability treating 
the back reaction in a systematic way was discussed 
in~\cite{Lewandowski:2001qp}.

Consider now the theory
\begin{eqnarray} \label{gwaction}
S &=& S_{bulk} + S_{vis} + S_{hid} \\
S_{bulk} & = &  \int d^4 x dz \sqrt{G}
\left(\frac{1}{2}(G^{MN} \partial_M \chi \partial_N \chi) - \frac{1}{2}
m^2 
\chi^2 \right) \nonumber\\
S_{vis} & = & - \int d^4 x \sqrt{g_{v}} \left( \lambda_{v1} \chi +
\frac{1}{2} 
\chi \lambda_{v2} \chi \right)  \nonumber\\
S_{hid} & = & - \int d^4 x \sqrt{g_h} \left( \lambda_{h1} \chi +
\frac{1}{2}\chi 
 \lambda_{h2} \chi \right),\nonumber
\end{eqnarray}
where we now use $k=1$ units. Note that $\lambda_{v1}, \lambda_{v2}$ are 
constants, not functions of $q$ like $\lambda_{v2}, \lambda_{h2}$.

The field $\chi$ must satisfy the boundary condition
\begin{equation}
z \partial_z \chi_q(z) \big|_{z=z_h}= \frac{1}{2}( \lambda_{h1} + 
\lambda_{h2}(qz_h) \chi_q(z_h) ).
\end{equation}
As before we attempt to match onto a theory with the new hidden 
boundary, 
\begin{equation} \label{gwbc}
z\partial_z \chi_q(z) \big|_{z=a}= \frac{1}{2}( \lambda_{1}(a) + 
\lambda_2(qa,a) 
\chi_q(a) ).
\end{equation}

As in the previous section, we take the logarithmic derivative of this 
boundary condition,
(\ref{gwbc}), and use the bulk equation of motion as well as 
 (\ref{gwbc})  to eliminate any $z$-derivatives of $\chi$.  
We arrive at the equation  
\begin{equation}~\label{tadandquad}
\Big(- \dota \lmb_1 + 4 \lmb_1 - \frac{1}{2} \lmb_1 \lmb_2 \Big)+\Big(-
\dota 
\lmb_2+ 4 \lmb_2 -\frac{1}{2} \lmb_2^2 + 2m^2 -2q^2 \Big) \chi_q(a)=0.
\end{equation}

We will formally treat the hidden boundary value of the field,
$\chi_q(a)$, as 
allowed to take arbitrary values, leading to RG equations
\begin{eqnarray}
 a \frac{\partial}{\partial a} \lambda_1 & = & 4 \lambda_1 - 
\frac{1}{2} \lambda_1 \lambda_2 \\
 a \frac{\partial}{\partial a} \lambda_2 & = & -qa \frac{\partial}{\partial 
(qa)} \lambda_2
+ 4 
\lambda_2 -\frac{1}{2} \lambda_2^2 + 
2m^2 -2q^2,
\end{eqnarray}
where the partial derivatives on the left act on the $a$ argument, not
on the 
$qa$ argument, which appears on the right. Clearly solutions to
these 
RG equations will solve (\ref{tadandquad}). However it is not actually
true 
that $\chi_q(a)$ can take on arbitrary values because of the extra
boundary 
conditions imposed by the visible brane. Thus we can view our RG
equations 
as the unique ones which are independent of the details of the visible
brane, 
very much like the more familiar
 mass and vev independent RG flows in energy scales.

 
In the Goldberger-Wise mechanism the scalar field sets up a stabilizing 
potential for the radion field the vev of which is $r_c \sim
\ln(z_v/z_h)$.  
Ignoring the gravitational backreaction as we have done gives the
leading 
approximation in Newton's constant  to the radion potential.
We may calculate this potential in the theory defined by
(\ref{gwaction}) as 
is 
done in Ref.~\cite{Lewandowski:2001qp}.  We may alternatively study the potential of the
equivalent 
effective theory with the hidden brane run down to
 a location $\sim z_v$ near the visible 
brane.   The couplings on the hidden brane will be 
modified from the theory with hidden brane at $z_h$ as required by the
RG 
flow. 

We consider a theory with hidden brane couplings close to the fixed
point.  We 
use the linearized flow equations.
\begin{eqnarray}
a \frac{\partial \lambda_1}{\partial a} & = & (4-\frac{\lambda_2^{*(0)}}{2})
\lambda_1 = 
(2-\nu)\lambda_1 \\
a \frac{\partial}{\partial a} (\lambda_2^{(0)}-\lambda_2^{*(0)}) & = & 
(4-\lambda_2^{*(0)})(\lambda_2^{(0)}-\lambda_2^{*(0)}) = -2 \nu 
(\lambda_2^{(0)} 
- \lambda_2^{*(0)})
\end{eqnarray} 
There are flow equations for the coefficients of higher derivative
operators, 
$\lambda_2^{(j>0)}$.  We do not consider these terms as we wish to find
the 
effective radion potential which is evaluated at zero external
momentum.  
Solving the flow equations we have
\begin{eqnarray}
\lambda_1(\sim 
z_v) & = & \lambda_{1h} \left(\frac{z_v}{z_h} \right)^{2-\nu} \\
\lambda^{(0)}_2(\sim z_v) & = & \lambda_2^{*(0)} + 
(\lambda^{(0)}_{2h}-\lambda_{2}^{*(0)}) \left(\frac{z_v}{z_h}
\right)^{-2\nu}
\end{eqnarray}
For such a small effective extra dimension we can neglect the bulk
action and 
treat the field $\chi$ as constant across the extra dimension. 
The effective action then becomes
\begin{equation}
 S = S_{a}+S_{vis} = \int\frac{d^4x}{z_v^4} \left( \left(\lambda_{v1} +
\lambda_{h1} 
\left(\frac{z_v}{z_h} \right)^{2-\nu} \right) \chi + \left( 
\lambda_{v2}^{(0)} + \lambda_{h2}^{*(0)} + (\lambda^{ (0)}_{h2}-\lambda_{2}^{*(0)}) 
\left(\frac{z_v}{z_h} \right)^{-2\nu} \right) 
\frac{\chi^2}{2} \right).
\end{equation}

As was found in Refs.~\cite{Goldberger:1999wh, Lewandowski:2001qp} a large hierarchy can be generated by a
moderately 
small scalar mass.  In this case we will find after radion stabilization
that 
$z_h/z_v \ll 1$.  We have
\begin{equation}
 2-\nu \sim -m^2/8, \qquad 2 \nu \sim 4.
\end{equation}
Therefore, contributions in the effective potential 
due to the deviation from $\lambda_2^*$ 
will be suppressed by $(z_h/z_v)^4$.  We may self-consistently neglect
this 
term.

The tadpole term in the action indicates a vev for $\chi$.  We find
\begin{equation}~\label{chivev}
<\chi> = \frac{\lambda_{v1} + \lambda_{h1}
(z_v/z_h)^{-m^2/8}}{\lambda_{v2}^{(0)} + 
\lambda_2^{*(0)}}.
\end{equation}
Note that this corresponds precisely to the solution for $\chi$ found in
Ref.~\cite{Lewandowski:2001qp}. There, a solution for $\chi$ to first order 
in $\lambda_{v1}$ and
$\lambda_{h1}$ 
was 
found 
as a function of the extra-dimensional space.  Evaluating this solution
at the 
location of the visible brane gives equation (\ref{chivev}).

Substituting this expression for $\langle \chi \rangle$ into the action
we
 obtain an effective potential for the radion.
\begin{equation}~\label{potential}
V_{eff} = -\frac{ \left( \lambda_{v1} + \lambda_{h1}
\left(\frac{z_v}{z_h} 
\right)^{- m^2/8} \right)^2}
{2 z_v^4 (\lambda_{v2}^{(0)} + \lambda_2^{*(0)})}.
\end{equation}
Minimizing this potential we find that a hierarchy is established.
\begin{equation}
\frac{z_v}{z_h} = \left( -\frac{\lambda_{v1}}{\lambda_{h1}}
\right)^{-8/m^2}.
\end{equation}
A large hierarchy is generated for a modest ratio of 
$\lambda_{v1}/\lambda_{h1}$.

\section{Interactions}
\label{interactions}

We now apply the methods of the previous sections to a 
more general action.  
Consider interacting theories of the type
\begin{eqnarray}~\label{action}
S_{bulk}[\chi(x,z)] & = & \iiv dz \sqrt{G} \Big\{ \frac{1}{2} G^{MN} 
\partial_M \chi \partial_N
\chi - V_{bulk}\big( \chi \big) \Big\} \nonumber
\\
S_{hid}[\chi(x,z_h)] & = & \iiv \sqrt{g_h} \; L_h( \chi, \partial_\mu, 
g_h^{\mu \nu})  \nonumber
\\
S_{vis}[\chi(x,z_v)] & = & \iiv \sqrt{g_v} \; L_v( \chi,\partial_\mu, 
g_v^{\mu \nu} ).
\end{eqnarray}
The brane localized 
Lagrangians are taken to be local in $\chi$ and its $x$-derivatives. We will
see 
that the RG flow preserves this general form.

The equations of motion
imply the hidden boundary condition
\be
z \partial_z \chi(x,z) \at{z=z_h} = \frac{-1}{2} \frac{1}{\sqrt{g_h}} 
\var{S_{hid}[\psi]}{\psi(x)} \at{\psi(x)=\chi(x,z_h)}.
\ee
We will use the dummy variable $\psi$ in the four dimensional variations
that 
remain after integrating over the delta functions in $z$ 
in the equations of motion emerging from brane terms.

In analogy to (\ref{bca}), we look for a new local action $S_a$ which 
satisfies 
the boundary condition
\be \label{genbca}
z \partial_z \chi(x,z) \at{z=a} =  \frac{-1}{2} \ginv 
\var{S_a[\psi]}{\psi(x)} \at{\psi(x)=\chi(x,a)}  \qquad a \in 
[z_h,z_v), 
\ee
where
\be
S_a[\psi] = \iiv \sqrt{g} \; L_a( a, \psi, \partial_\mu, g^{\mu \nu}),
\qquad 
g^{\mu \nu}= G^{\mu \nu}(x,z=a).
\ee
We require the boundary condition
\be
S_a[\chi(x,z_h)]  = S_{hid}[\chi(x,z_h)],
\ee
and that solutions, $\chi(x,z)$, 
to the original equations of motion  solve the effective 
equations of motion  when restricted to $z \in [a,z_v)$.

As in previous sections, we 
 calculate the logarithmic $a$-derivative of the effective  hidden 
boundary condition, and eliminate first and second $z$-derivatives at
the 
effective hidden brane using the boundary condition and the equations of 
motion. Using the obvious result $a \partial_a g^{\mu \nu} = 2 g^{\mu
\nu}$ 
we arrive at
\begin{eqnarray} \label{chillipepper}
\frac{1}{2 \sqrt{g}} a \partial_a \frac{\delta S_a[\psi]}{\delta 
\psi(x)}\bigg|_{\psi(x)=\chi(x,a)} 
=  
\frac{1}{4 (\sqrt{g})^2} \int d^4 x^{\prime} \frac{\delta^2 
S_a[\psi]}{\delta \psi(x^{\prime}) \delta \psi(x) 
}\bigg|_{\psi(x)=\chi(x,a)} \times \frac{\delta S_a[\psi]}{\delta
\psi(y)} 
\bigg|_{\psi(y) = \chi(x^{\prime},a)} && \nonumber \\
-
\frac{1}{\sqrt{g}} \int d^4 x^{\prime \prime} \frac{\delta^2 
S_a[\psi(x')]}{\delta g^{\mu \nu}(x^{\prime \prime}) \delta \psi(x) 
}\bigg|_{\psi(x)=\chi(x,a)}  g^{\mu \nu}(x^{\prime \prime}) 
-
\frac{\partial V_{bulk}}{\partial \chi(x,a)} - g^{\mu \nu}
\partial_{\mu} 
\partial_{\nu} \chi(x,a), &&
\end{eqnarray}
where the partial $a$ derivative on the left is taken to act only on the 
running couplings of the effective theory.

It is important to note that this
 equation is local in the couplings and the field $\chi(x,a)$ since it
is 
constructed from 
the variations of the presumed local action, $S_a$. Thus we have shown
that 
local effective actions flow to local effective actions, and the
boundary 
condition for the flow at $a = z_h$ is $S_{hid}$, which is local.
 To find the $\beta$ functions for each coupling, we expand 
(\ref{chillipepper}) in powers of $\chi_q(a)$ and the independent
momenta, $q$
 and require that each coefficient vanish. As in the previous section,
this 
amounts to treating $\chi_q(a)$ as free to take on any value in the 
set of 5D solutions, which is equivalent to choosing an RG flow which is 
independent of the details of the visible brane.

Since we have not considered higher-derivative interactions in the bulk, 
we have still not obtained the greatest generality. There is no
obstruction 
to finding local RG flows in the most general case of higher bulk
derivatives,
an example of which appears in Section \ref{bulkderivatives}.


As an example of the use of equation (\ref{chillipepper}), 
take the theory with 
the bulk interaction
\begin{equation}
S_{bulk} \supset \int d^4 x dz \sqrt{G} (\sigma_3 \chi^3).
\end{equation}
On the hidden brane we have
\begin{equation}
S_{hid} \supset \int d^4 x \sqrt{g_h} (\lambda_{h3}^{(0)} \chi^3 + 
\lambda_{h4}^{(0)} \chi^4).
\end{equation}  
   We can calculate the running of $\lambda_3^{(0)}$ and
$\lambda_4^{(0)}$.  
Equation (\ref{chillipepper}) gives
\begin{eqnarray}
&& \dots + 3 (a \frac{\partial}{\partial a} \lambda_3^{(0)} -4
\lambda_3^{(0)} + 
\frac{3}{2} \lambda_2^{(0)} \lambda_4^{(0)} + 2 \sigma_3^{(0)}) \chi^2 
\nonumber 
\\
&& \qquad + 4 (a \frac{\partial}{\partial a} \lambda_4^{(0)} -4
\lambda_4^{(0)}
 + 
2 \lambda_2^{(0)} \lambda_4^{(0)} + \frac{9}{4}
(\lambda_3^{(0)})^2)\chi^3  + 
\ldots  = 0
\end{eqnarray}
where we take $\lambda_{h1} = 0$.  Our RG flow then 
follows 
by treating $\chi(z = a)$ as a free variable,
\begin{eqnarray}
a \partial_a \lambda_3^{(0)} & = & 4 \lambda_3^{(0)} - \frac{3}{2} 
\lambda_2^{(0)} \lambda_4^{(0)} - 2 \sigma_3^{(0)} \\
a \partial_a \lambda_4^{(0)} & = & 4 \lambda_4^{(0)} - 2 \lambda_2^{(0)} 
\lambda_4^{(0)} - \frac{9}{4} (\lambda_3^{(0)})^2.
\end{eqnarray}

A central result of this present section is  the robust retention of
a
 fixed point in the RG flow in the
presence of interactions, which is
 IR-attractive as long as the bulk mass-squared
is 
positive.  To see 
this recall that we have found in Section II a fixed point for the free  
theory:
\begin{eqnarray}~\label{freefp}
\lambda_2 & = & \lambda_2^* \\
\lambda_n & = & 0
\end{eqnarray}
Using (\ref{chillipepper}) one can straightforwardly 
find the linearized $\beta$ functions
for 
flows near this fixed point when brane interactions are permitted, but
bulk 
interactions are still absent.  We find
\begin{equation}
a \frac{\partial}{\partial a}\vec{\lambda} = {\bf \gamma} \cdot
(\vec{\lambda} - 
\vec{\lambda}^*) + {\cal O}((\vec{\lambda} - \vec{\lambda}^*)^2).
\end{equation}
The scaling matrix, ${\bf \gamma}$, is lower triangular (operators with 
derivatives only mix with operators with fewer derivatives and
$\lambda_n$ 
and 
$\lambda_m$ do not mix for $n \neq m$), so that the eigenvalues are given by 
the diagonal elements.  These eigenvalues are
\begin{equation}
\gamma_i^{(2j)} = 4-2j-\frac{i}{2}(4+2 \nu) < 0
\end{equation}
where $i$ indicates the number of $\chi$ fields in the operator and $2j$ 
indicates the highest number of derivatives in that operator.  The
scaling 
dimensions are negative so this fixed point is attractive.

We  now add bulk interactions. 
  From (\ref{chillipepper}) one can see that the $\beta$ function 
will be corrected by a power series in bulk couplings (actually a linear 
term at classical order).  
 If these couplings are given the formal small parameter $\alpha$ they
will 
alter the RG flow by ${\cal O}(\alpha)$ terms, which cannot eliminate
the fixed
point but only shift its position by  ${\cal O}(\alpha)$.   The scaling 
dimensions will also receive an ${\cal O}(\alpha)$ correction.  The new fixed
point 
will therefore be attractive for a sufficiently small $\alpha$.

\section{Goldberger-Wise mechanism in the interacting theory}

The robustness of the fixed point and the anomalous dimensions when 
interactions are included translates into 
the robustness of the Goldberger-Wise mechanism for stabilizing the 
hierarchy, which we see as follows. As long as our original hidden brane 
action is sufficiently close to the fixed point action, we are justified 
in using the linearized RG flow, given by the anomalous dimensions. As
long as
the strength of bulk couplings is sufficiently small, the $\alpha$
mentioned 
above being at most of order the bulk mass-squared (in $k = 1$ units),
the 
spectrum of anomalous dimensions is still given by one coupling with
anomalous 
dimension of order $- m^2$, while the other couplings have anomalous
dimensions
which are of order $-1$ or more negative. 

We will now repeat our 
Goldberger-Wise analysis as in the previous section, but now using this 
interacting RG flow. 
Running the Planck brane down to $z_v$ as before we drop derivative terms 
to find a vacuum 
potential of the form
\begin{equation}
V_{eff} = \frac{1}{z_v^4} \left( L_v(\chi) + L^*_h(\chi)+ 
\sum_i (\lambda_{hi} -
 \lambda_{hi}^*)\left(\frac{z_v}{z_h} \right)^{\gamma_i} {\cal O}_i(\chi) 
\right)
\end{equation}
where the operators ${\cal O}_i$ have scaling dimension $\gamma_i$. The functions $L_v$ and $L_h^*$ have no explicit $z_v/z_h$ dependence. We will 
show self-consistently that $z_h/z_v \ll 1$.  We can then
discard all 
terms suppressed by order one powers of $z_h/z_v$ as they are negligible. 
 Taking 
$\gamma_1$ to be of order $-m^2$, only one hidden brane operator is relevant.
We have
\begin{equation}
V_{eff} = \frac{1}{z_v^4} \left( L_v(\chi) + L^*_h(\chi)+  (\lambda_{h1} -
 \lambda_{h1}^*)\left(\frac{z_h}{z_v} \right)^{\gamma_1} {\cal O}_1(\chi) 
\right).
\end{equation}
Notice that this potential has the form,
\begin{equation}
V_{eff} = \frac{1}{z_v^4} P(\chi, (z_v/z_h)^{\gamma_1}),
\end{equation}
where $P$ indicates a polynomial.  For quite generic coefficients 
this indicates that solutions with
\begin{eqnarray}
\langle \chi \rangle & = & {\cal O}(1), \\
\left(\frac{z_v}{z_h} \right)^{\gamma_1} & = & {\cal O}(1),
\end{eqnarray}
minimize this potential.
 These relations are explicitly confirmed at leading order 
in $\alpha$ in Section III were a large hierarchy is generated.  We see that a
 large 
hierarchy will also be set including the ${\cal O}(\alpha)$ (and higher)
corrections.
\begin{equation}
\frac{z_v}{z_h} = \left( {\cal O}(1) \right)^{1/\gamma_1}.
\end{equation}

\section{Green's Functions}

In quantum field theory the quantities of physical interest are the
$n$-point 
functions.  Take the theory of equation (\ref{action}) with the hidden brane at
$z=z_h$.  
We may also consider a theory with the hidden brane at a location
$a>z_h$.  
These theories will describe equivalent physics if they give the same 
$n$-point 
functions.  We require
\begin{equation}
\langle \chi(x_1,z_1) \ldots \chi(x_n,z_n) \rangle_{z_h} =  \langle 
\chi(x_1,z_1) \ldots \chi(x_n,z_n) \rangle_{a}
\end{equation}
where the final subscript denotes the position of the  hidden
brane,  where
 $z_i > a$.  This method of matching $n$-point functions is done at tree 
level in this paper and gives an alternate, equivalent way of calculating 
the 
flow 
equations for hidden brane couplings to that of the previous sections. 
We 
demonstrate this method through a few examples.  Its utility is
primarily in 
that the method may be extended to quantum calculations.

\subsection{The 2 point function}

To evaluate $n$-point functions in perturbation theory we must first
consider 
 the two-point function.  
\begin{equation}
 \Delta(x,z;x',z') \equiv \langle \chi(x,z) \chi(x',z') \rangle
\end{equation}
It satisfies
\begin{equation}~\label{greeneom}
\left( \partial_M \sqrt{G} G^{MN} \partial_N + \sqrt{G} m^2 + \sqrt{g_v} 
\lambda_{v2} \delta(z-z_v) + \sqrt{g_h} \lambda_{h2} \delta(z-z_h) \right) 
\Delta(x,z;x',z') = \delta^4(x-x') \delta(z-z')
\end{equation}
Explicit solutions for the Green's function in AdS space have been found in 
Refs.~\cite{Grinstein:2000ny,Giddings:2000mu}.
Simplifying in 4D momentum space this is
\begin{equation}
\frac{1}{z^3} \left( -q^2 - \partial_z^2 + \frac{3}{z}\partial_z +
 \frac{m^2}{z^2} + \frac{\lambda_{v2}}{z_v} \delta(z-z_v) +
\frac{\lambda_{h2}}{z_h} 
\delta(z-z_h) \right) \Delta_q(z,z')= \delta(z-z').
\end{equation}
As usual, this is equivalent to a  $z_h$ independent differential
equation
\begin{equation}~\label{gfbulk}
\frac{1}{z^3} \left( -q^2 - \partial_z^2 + \frac{3}{z}\partial_z + 
\frac{m^2}{z^2} \right) \Delta_q(z,z') = \delta(z-z'),
\end{equation}
subject to  $z_h$-dependent boundary conditions, 
\begin{eqnarray}
\partial_z \Delta_q (z,z') \bigg|_{z=z_h} & = &
\frac{\lambda_{2h}(qz_h)}{2  z_h} 
\Delta_q(z_h,z')  \\
\partial_z \Delta_q (z,z') \bigg|_{z=z_v} & = & - \frac{\lambda_{v2}}{2  z_v} 
\Delta_q(z_v,z'). 
\end{eqnarray}
In the effective theory with the hidden brane at $a$ the 2-point
function
satisfies the same differential equation and visible boundary condition
but 
with a modified hidden boundary
condition.
\begin{equation}
\partial_z \Delta_q(z,z') \bigg|_{z=a}  =  \frac{\lambda_2 (qa,a)}{2  a} 
\Delta_q(a,z'). 
\end{equation}
The coupling $\lambda_2(qa,a)$ is chosen so that 
\begin{equation}
\Delta_q(z,z')_a = \Delta_q(z,z')_{z_h}, \qquad z,z' > a,
\end{equation}
which implies
\begin{equation} \label{greenbc}
\partial_z \Delta_q(z,z')_a \bigg|_{z=a+\delta a}  =  \frac{\lambda_2 
(q(a+\delta a), a+\delta a)}{2  (a+\delta a)} \Delta_q(a+\delta a,z')_a. 
\end{equation}
Expanding for infinitesimal $\delta a$, using (\ref{gfbulk}) to
eliminate 
$\partial_z^2 \Delta$  and the effective hidden boundary condition to 
eliminate $\partial_z \Delta$ at $z = a$, leads to 
\begin{equation}~\label{lambda2beta}
\sum_{j} (qa)^j a \frac{\partial \lambda_2^{(j)}}{\partial a}  = - qa 
\frac{\partial \lambda_2}{\partial qa} + 4 \lambda_2
-\frac{\lambda_2^2}{2}+ 2 
m^2 - 2 (qa)^2.
\end{equation}
This is the same flow described in Section \ref{free}.  

What follows are two example flow calculations.  The central object will
be 
the free field two 
point function at the RG fixed point, satisfying the boundary condition
\begin{equation}
\partial_z \Delta_q(z,z') \bigg|_{z=a}  =  \frac{\lambda_2^* (qa)}{2  a} 
\Delta_q(a,z'). 
\end{equation}
The theory will in general not be at this fixed point.  We treat
contributions 
to the $n$-point function due to bulk interactions and 
deviations from $\lambda_2^*$ perturbatively.

\subsection{Bulk Derivatives}
\label{bulkderivatives}

In our earlier classical analysis we had restricted ourselves for simplicity 
to bulk interactions of the form of a potential. 
As an example of a perturbative calculation using the method of matching 
Green's 
functions consider here the specific 
bulk derivative coupling  
\begin{equation}
S_{bulk} \supset \int d^4 x \int dz \sqrt{G} \sigma_4^{(2)} (
G^{MN} 
\partial_M  \chi  \partial_N \chi) (\chi^2), 
\end{equation}
with coupling constant $\sigma_4^{(2)}$~\footnote[1]{
We will abstain from doing the most general analysis of bulk derivative 
couplings
due to the sheer complexity of notation rather than any intrinsic difficulty. 
Our example will allow the reader to generalize to any case of their interest.
Note that couplings involving more than one $z$-derivative acting on the
field can 
be removed by a field redefinition as
described in 
Ref. \cite{Lewandowski:2001qp}.}
Take two theories with the hidden brane at $a$ and $a+\delta a$
respectively.  
We require that the 4-point function be independent of the location of
the 
brane,
\begin{equation}
\langle \chi(x_1,z_1) \ldots \chi(x_4,z_4) \rangle_{a} = \langle
\chi(x_1,z_1) 
\ldots \chi(x_4,z_4) \rangle_{a+\delta a}.
\end{equation}
For infinitesimal $\delta a$ this is
\begin{equation}
a \frac{d}{da} \langle \chi(x_1,z_1) \ldots \chi(x_4,z_4) \rangle_{a} =
0.
\end{equation}
On the hidden brane we need
\begin{equation}
S_{a} \supset \int d^4 x \sqrt{g_a} (\lambda_4^{(0)} \chi^4 +
\ldots ).
\end{equation}
where in this section $\ldots$ indicates terms with higher 
$x^{\mu}$ derivatives.  In 
perturbation theory we use the free field propagator at the fixed point 
$\lambda_2^*$, as discussed in the previous subsection.

Calculating the 4 point function at first order in the couplings gives 
\begin{eqnarray}
&& \langle \chi(x_1,z_1) \ldots \chi(x_4,z_4) \rangle_{a} = \nonumber \\
&& 4 \int d^4 x \int_{z_v}^{a} dz \sigma_4^{(2)} \sqrt{G} G^{zz} \left( 
\partial_z \Delta(x_1,z_1;x,z) \partial_z \Delta(x_2,z_2;x,z) 
\Delta(x_3,z_3;x,z) \Delta(x_4,z_4;x,z) + {\rm perm.} \right) \nonumber
\\ 
&&+ 12 \int d^4 x \sqrt{ g_a} \lambda_4^{(0)} \Delta(x_1,z_1;x,a) 
\Delta(x_2,z_2;x,a) \Delta(x_3, z_3;x,a) \Delta(x_4,z_4;x,a) + \ldots
\end{eqnarray}
where ${\rm perm.}$ indicates a permutation of the $z$ dervatives on the 
propagators.  Taking the logarithmic derivative with respect to $a$ we
have
\begin{eqnarray}
&& 4 \int d^4 x  \Bigg\{ -\frac{1}{a^4} \sigma_4^{(2)} a^2  \left( \partial_a 
\Delta(x_1,z_1;x,a) \partial_a \Delta(x_2,z_2;x,a) \Delta(x_3,z_3;x,a) 
\Delta(x_4,z_4;x,a) + {\rm perm.} \right) \nonumber \\
&& + 12 \bigg( \frac{4}{a^4} \lambda_4^{(0)} \Delta(x_1,z_1;x,a) 
\Delta(x_2,z_2;x,a) \Delta(x_3, z_3;x,a) \Delta(x_4,z_4;x,a) \nonumber
\\
&& 
 -\frac{1}{a^4} \left(a \frac{d}{da}\lambda_4^{(0)} \right)  
\Delta(x_1,z_1;x,a) \Delta(x_2,z_2;x,a) \Delta(x_3, z_3;x,a) 
\Delta(x_4,z_4;x,a)  \nonumber \\
&& -\frac{a}{a^4} \lambda_4^{(0)} \frac{d}{da} \left(
\Delta(x_1,z_1;x,a) 
\Delta(x_2,z_2;x,a) \Delta(x_3, z_3;x,a) \Delta(x_4,z_4;x,a) \right) 
\bigg) + 
\ldots \Bigg\} =0.
\end{eqnarray}
Using the fact that the two-point functions are at their free-field
fixed point values,  
\begin{equation}
 a \frac{d}{da} \Delta(x,a;x_i,z_i)  =  \frac{\lambda_2^*}{2} 
\Delta(x,a;x_i,z_i),
\end{equation}
we obtain
\begin{eqnarray}
&&\frac{12}{a^4} \int d^4 x \Delta(x_1,z_1;x,a) \Delta(x_2,z_2;x,a) 
\Delta(x_3, 
z_3;x,a) \Delta(x_4,z_4;x,a) \times \nonumber \\
&& \qquad \bigg(2 \sigma_4^{(2)} a^2 \left( \frac{\lambda_2^{*(0)}}{2a} 
\right)^2 -4 \lambda_4^{(0)}+ a \frac{d}{da} \lambda_4^{(0)} +  4 a 
\lambda_4^{(0)} \left(\frac{\lambda_2^{*(0)}}{2a}\right) \bigg) + \ldots
=0
\end{eqnarray}
In 4-momentum space we may divide by the external propagators and then set 
external 
momentum to zero to obtain
\begin{equation}
a \frac{d 
\lambda_4^{(0)}}{d a} = 
(4- 2 \lambda_2^{*(0)}) \lambda_4^{(0)} - 2 \sigma_4^{(2)} \left( 
\frac{\lambda_2^{*(0)}}{2}\right)^2,
\end{equation}
showing how this bulk derivative interaction modifies the RG flow for
a hidden brane coupling.

\subsection{Nonlinear effects of brane $\chi^3$ interactions}

As a second example, illustrating nonlinear effects in the RG, consider
 the brane interactions
\begin{equation}
S_{hid} \supset \int d^4 x \sqrt{g_h} \lambda_{h3}^{(0)} \chi^{3}.
\end{equation} 

\begin{figure} 
   \epsfxsize=4in 
   \centerline{\epsfbox{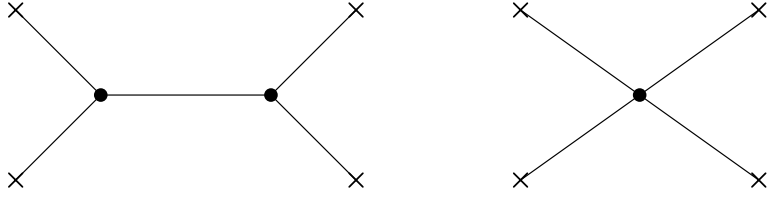}} 
   \caption{Diagrams giving the four point function at tree level.  A $\chi^3$ interaction requires a $\chi^4$ ``counterterm'' at tree level to match theories with different radii.} 
   \label{fourpoint} 
\end{figure}

These will induce 4-point operators through the diagrams of 
Figure \ref{fourpoint}.  
We require hidden brane ``counterterms''
\begin{equation}
S_{{\rm c.t.}} \supset \int d^4 x \sqrt{g_h} (\lambda_4^{(0)} \chi^4 + \ldots).
\end{equation}
 to maintain invariance of the 4-point function working to second order in $\lambda_{h3}^{(0)}$. 
Here again
$\ldots$ indicates higher $x_{\mu}$ derivatives.

\begin{figure} 
   \epsfxsize=4in 
   \centerline{\epsfbox{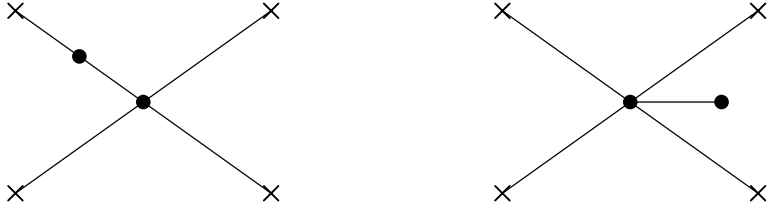}} 
   \caption{Diagrams contributing to the fourpoint function in addition to 
those of Figure 1.  They give nonlinear contributions to the $\beta$ function 
of the form $\lambda_2 \lambda_4$ (left fig.) and 
$\lambda_1 \lambda_5$ (right fig.).} 
   \label{fourint} 
\end{figure}

We calculate
\begin{eqnarray}~\label{fourpointcalc}
&& \langle \chi(x_1,z_1) \ldots \chi(x_4,z_4) \rangle_{a} = \nonumber \\
&&  12 \int d^4 x \frac{\lambda_4^{(0)}}{(a)^4} \Delta(x_1,z_1;x,a) 
\Delta(x_2,z_2;x,a) \Delta(x_3, z_3;x,a) \Delta(x_4,z_4;x,a) \nonumber
\\
&&
+ 9 \int d^4 x d^4 y \frac{(\lambda_3^{(0)})^2}{(a)^8} \bigg( 
\Delta(x_1,z_1;x,a) \Delta(x_2,z_2;x,a) \Delta(x,a;y,a)  \Delta(x_3,
z_3;y,a) 
\Delta(x_4,z_4;y,a)
   \nonumber  \\
&&
+ (x_2,z_2 \leftrightarrow x_3,z_3) + (x_2,z_2 \leftrightarrow x_4,z_4)
\bigg) + 
\ldots  
\end{eqnarray}
requiring
\begin{equation}
a \frac{d}{da} \langle \chi(x_1,z_1) \ldots \chi(x_4,z_4) \rangle_{a} =
0.
\end{equation}
We use
\begin{equation}
 a \frac{d}{da} \Delta(x,a;x_i,z_i)  =  \frac{\lambda_2^*}{2} 
\Delta(x,a;x_i,z_i),
\end{equation}
\begin{equation}~\label{disc}
a \frac{d}{da} \Delta(x,a;y,a) = \lambda_2^* 
\Delta(x,a,y,a) - a^4 \delta^4 (x-y)
\end{equation}
where the delta function term in (\ref{disc}) is due to the discontinuity
in $\Delta$ following from (\ref{greeneom}).
To linear order in $\lambda_3^{(0)}$ it is straightforward to establish
\begin{equation}
a \frac{d}{da} \lambda_3^{(0)} = (4-\frac{3}{2} 
\lambda_2^{*(0)}) \lambda_3^{(0)}.
\end{equation}
Substituting these expressions into (\ref{fourpointcalc}), transforming to 
4-momentum space, dividing by 
the external propagators and setting external momentum to zero
gives
\begin{equation}
a \frac{d}{da} \lambda_4^{(0)} = (4-2 \lambda_2^{*(0)}) \lambda_4^{(0)}
- 
\frac{9}{4} (\lambda_3^{(0)})^2
\end{equation}
This matches the result that is obtained through the methods of 
Section \ref{interactions}.  This 
demonstrates how $\chi^3$ interactions induce running in 
$\lambda_4^{(0)}$.  Note that other nonlinear contributions in the $\beta$
functions may 
be calculated through the diagrams of Figure \ref{fourint}.

\section{Concluding Remarks}

The two main directions for generalizing the above RG formalism are  
including higher spins and including  quantum loops. Spins greater than or 
equal to one raise an interesting subtlety not encountered in this paper. 
The usual CFT interpretation of a bulk RS1 field is that it corresponds to 
a CFT operator {\it and} to a fundamental 4D field coupled to that CFT. 
Therefore it is at first sight odd that in our RS1 examples we
found exact fixed points without the deviations corresponding
to the coupling to a fundamental 4D scalar.  However, one 
can see that the hidden brane fixed point Lagrangian 
found in this paper contains an effective-cutoff sized mass for the bulk 
scalar. The dual of this is a CFT coupled to a cutoff-mass fundamental 
scalar that is a scalar heavy enough to be completely integrated out of 
the effective theory leaving the pure CFT dynamics. 
Now consider higher spins protected by 
gauge invariance.  For concreteness we will consider gravity in the full 
RS1 model.   
We may want to keep the zero mode graviton (dual to the fundamental 
4D graviton coupled to 
the CFT).  In this case we must accomodate 
 the complication that 
the effective hidden brane theory must describe ``almost'' fixed point 
behavior with small deviations arising from weak coupling to the zero mode
graviton. 

Quantum loops can be dealt with in a very similar manner to the methods of 
Section VI. The central technical issue is a proof to all orders of the 
locality of the running effective hidden brane Lagrangian. (We fully expect 
this locality.) Once this is demonstrated, the RG approach will 
undoubtedly provide the simplest and most insightful proof of the stability 
of the RS-Goldberger-Wise hierarchy to all orders in the effective field 
theory. The method of proof would be almost identical to the proof of 
hierarchy robustness given here in Section V for the classical interacting
theory, which uses only the most general features of the 
RG flows. The importance of such a proof is easy to understand when one 
considers that the Goldberger-Wise effective potential is basically a 
weak-scale potential for the radius, whereas the RS1 effective theory, 
has access to Planckian mass scales (especially at the quantum level) which 
naively threaten to overwhelm weak-scale effects. Understanding how the
 potential is protected by such effects is
a particular case illustrating 
 the more general utility of the RG procedure, which is 
to identify where large dependences on the warp factor appear 
within Feynman diagrams. Another important application would be quantum 
corrections to the masses and decays of low lying Kaluza-Klein resonances.

Work on these fronts is in progress.

\acknowledgments
The research of A.\ L., M.\ M. and R.\ S. is supported by 
NSF Grant P420D3620434350.
R.S. is also supported by a DOE Outstanding Junior Investigator award 
Grant P442D3620444350.


\begin{thebibliography}{99}

\bibitem{Randall:1999ee}
L.~Randall and R.~Sundrum,
Phys.\ Rev.\ Lett.\  {\bf 83}, 3370 (1999)
[arXiv:hep-ph/9905221];

\bibitem{Randall:1999vf}
L.~Randall and R.~Sundrum,
Phys.\ Rev.\ Lett.\  {\bf 83}, 4690 (1999)
[arXiv:hep-ph/9906064].

\bibitem{adshard}
W.~D.~Goldberger and I.~Z.~Rothstein,
Phys.\ Lett.\ B {\bf 491}, 339 (2000)
[arXiv:hep-th/0007065];
L.~Randall and M.~D.~Schwartz,
JHEP {\bf 0111}, 003 (2001)
[arXiv:hep-th/0108114];
W.~D.~Goldberger and I.~Z.~Rothstein,
[arXiv:hep-th/0204160];
K.~Agashe, A.~Delgado and R.~Sundrum,
[arXiv:hep-ph/0206099];
R.~Contino, P.~Creminelli and E.~Trincherini,
[arXiv:hep-th/0208002].
K.~w.~Choi and I.~W.~Kim,
[arXiv:hep-th/0208071].

\bibitem{holography}
J.~M.~Maldacena,
Adv.\ Theor.\ Math.\ Phys.\  {\bf 2}, 231 (1998)
[Int.\ J.\ Theor.\ Phys.\  {\bf 38}, 1113 (1999)]
[arXiv:hep-th/9711200];
S.~S.~Gubser, I.~R.~Klebanov and A.~M.~Polyakov,
Phys.\ Lett.\ B {\bf 428}, 105 (1998)
[arXiv:hep-th/9802109];
E.~Witten,
Adv.\ Theor.\ Math.\ Phys.\  {\bf 2}, 253 (1998)
[arXiv:hep-th/9802150];
L.~Susskind and E.~Witten,
[arXiv:hep-th/9805114];
E.~T.~Akhmedov,
[arXiv:hep-th/9806217];
E.~Alvarez and C.~Gomez,
Nucl.\ Phys.\ B {\bf 541}, 441 (1999)
[arXiv:hep-th/9807226];
V.~Balasubramanian and P.~Kraus,
Phys.\ Rev.\ Lett.\  {\bf 83}, 3605 (1999)
[arXiv:hep-th/9903190];
S.~Ferrara and M.~Porrati,
Phys.\ Lett.\ B {\bf 458}, 43 (1999)
[arXiv:hep-th/9903241];
D.~Z.~Freedman, S.~S.~Gubser, K.~Pilch and N.~P.~Warner,
JHEP {\bf 0007}, 038 (2000)
[arXiv:hep-th/9906194];



\bibitem{rsholography}
H. Verlinde, Nucl.\ Phys.\ B {\bf 580}, 264 (2000) [arXiv:hep-th/9906182];
J.~Maldacena, unpublished remarks; 
E.~Witten, ITP Santa Barbara conference `New Dimensions in Field Theory and String Theory', http://www.itp.ucsb.edu/online/susyc99/discussion;
S.~S.~Gubser,
Phys.\ Rev.\ D {\bf 63}, 084017 (2001)
[arXiv:hep-th/9912001];
J.~de Boer, E.~Verlinde and H.~Verlinde,
JHEP {\bf 0008}, 003 (2000)
[arXiv:hep-th/9912012];
N.~Arkani-Hamed, M.~Porrati and L.~Randall,
JHEP {\bf 0108}, 017 (2001)
[arXiv:hep-th/0012148];
R.~Rattazzi and A.~Zaffaroni,
JHEP {\bf 0104}, 021 (2001)
[arXiv:hep-th/0012248];
M.~Perez-Victoria,
JHEP {\bf 0105}, 064 (2001)
[arXiv:hep-th/0105048].
C.~Csaki, J.~Erlich, T.~J.~Hollowood and J.~Terning,
Phys.\ Rev.\ D {\bf 63}, 065019 (2001)
[arXiv:hep-th/0003076].


\bibitem{Goldberger:1999wh}
W.~D.~Goldberger and M.~B.~Wise,
Phys.\ Rev.\ D {\bf 60}, 107505 (1999)
[arXiv:hep-ph/9907218].


\bibitem{skenderis}
M.~Henningson and K.~Skenderis,
JHEP {\bf 9807}, 023 (1998)
[arXiv:hep-th/9806087].
M.~Henningson and K.~Skenderis,
Fortsch.\ Phys.\  {\bf 48}, 125 (2000)
[arXiv:hep-th/9812032].
S.~de Haro, S.~N.~Solodukhin and K.~Skenderis,
Commun.\ Math.\ Phys.\  {\bf 217}, 595 (2001)
[arXiv:hep-th/0002230].
M.~Bianchi, D.~Z.~Freedman and K.~Skenderis,
JHEP {\bf 0108}, 041 (2001)
[arXiv:hep-th/0105276].
M.~Bianchi, D.~Z.~Freedman and K.~Skenderis,
Nucl.\ Phys.\ B {\bf 631}, 159 (2002)
[arXiv:hep-th/0112119].
S.~de Haro, K.~Skenderis and S.~N.~Solodukhin,
Class.\ Quant.\ Grav.\  {\bf 18}, 3171 (2001)
[arXiv:hep-th/0011230].



\bibitem{wiseflow}
W.~D.~Goldberger and M.~B.~Wise,
Phys.\ Rev.\ D {\bf 65}, 025011 (2002)
[arXiv:hep-th/0104170].

\bibitem{Milton}
K.~A.~Milton, S.~D.~Odintsov and S.~Zerbini,
Phys.\ Rev.\ D {\bf 65}, 065012 (2002)
[arXiv:hep-th/0110051].


\bibitem{deconstruction}
L.~Randall, Y.~Shadmi and N.~Weiner,
[arXiv:hep-th/0208120].


\bibitem{falkowski}
A.~Falkowski and H.~D.~Kim,
JHEP {\bf 0208}, 052 (2002)
[arXiv:hep-ph/0208058].




\bibitem{DeWolfe:1999cp}
O.~DeWolfe, D.~Z.~Freedman, S.~S.~Gubser and A.~Karch,
Phys.\ Rev.\ D {\bf 62}, 046008 (2000)
[arXiv:hep-th/9909134].

\bibitem{Lewandowski:2001qp}
A.~Lewandowski and R.~Sundrum,
Phys.\ Rev.\ D {\bf 65}, 044003 (2002)
[arXiv:hep-th/0108025].

\bibitem{Grinstein:2000ny}
B.~Grinstein, D.~R.~Nolte and W.~Skiba,
Phys.\ Rev.\ D {\bf 63}, 105005 (2001)
[arXiv:hep-th/0012074].


\bibitem{Giddings:2000mu}
S.~B.~Giddings, E.~Katz and L.~Randall,
JHEP {\bf 0003}, 023 (2000)
[arXiv:hep-th/0002091].

\end{thebibliography}
\end{document}